\documentclass[a4paper,11pt]{article}

\usepackage{contribution}
\usepackage{dsfont}
\usepackage{epsfig}

\newcommand{\weblink}[2][]{%
    \ifthenelse{\equal{#1}{}}%
    {\textnormal{\url{#2}}}%
    {\textnormal{\href{#2}{#1}}}%
}

\def\beq{\begin{equation}}
\def\eeq#1{\label{#1}\end{equation}}
\def\eeqn{\end{equation}}

\def\beqa{\begin{eqnarray}}
\def\eeqa#1{\label{#1}\end{eqnarray}}
\def\eeqan{\end{eqnarray}}

\let\bar=\overbar

\def\Dslash{\not{\hbox{\kern-4pt $D$}}}
\def\dslash{\not{\hbox{\kern-2pt $\del$}}}

\def\msb{{\bar{\ssstyle M \kern -1pt S}}}

\newcommand{\contribution}[7][]{%
  \clearpage
  \thispagestyle{plain}
  \ifthenelse{\equal{#1}{}}
  {\hypersetup{pdftitle={#2}}}
  {\hypersetup{pdftitle={#1}}}
  \hypersetup{pdfauthor={{#3} {#4}}}
  {\centering\normalfont\LARGE\bfseries\sffamily #2 \par\nobreak}
  \lhead{}
  \chead{%
    \textit{\footnotesize XIV International Conference on Hadron Spectroscopy
      (\weblink[\textit{hadron2011}]{http://www.hadron2011.de}), 13-17 June 2011, Munich, Germany}%
  }
  \rhead{}
  \bigskip
  \begin{center}
    {#3} {#4}\ifthenelse{\equal{#6}{}}{}{\footnote{\weblink[#6]{mailto:#6}}}
    \ifthenelse{\equal{#7}{}}{}{#7} \\
    \textit{#5}
  \end{center}
  \bigskip
}

\renewcommand{\abstract}[1]{%
  \begin{center}
    \begin{minipage}{0.85\textwidth}
      \begin{footnotesize}
        #1
      \end{footnotesize}
    \end{minipage}
  \end{center}
  \bigskip
}

\begin{document}


%
%
%
%

{  


%
\contribution[The $K^- d \to \Lambda(1405)$ n reaction]
{The K d -> Lambda(1405) n reaction with the DAFNE set up and the $\bar{K}NN$ system revisited}
{E.}{Oset}  
{$^1$Departamento de F\'isica Te\'orica and IFIC, Centro Mixto\\ Universidad de
Valencia-CSIC, Institutos de Investigaci\'on de Paterna, Aptdo. 22085, 46071
Valencia, Spain\\
$^2$Yukawa Institute for Theoretical Physics, 
Kyoto University, Kyoto 606-8502, Japan} 
{}
{$^1$, D. Jido$^2$, T. Sekihara$^2$, M. Bayar$^1$ and J.Yamagata-Sekihara$^1$}



\abstract{The $K^{-}$ induced production of $\Lambda(1405)$ in the
$K^{-} d \to \pi \Sigma n$ reaction is investigated having in mind the conditions of the 
DAFNE facility at Frascati. We find that the fastest kaons from the decay of the $\phi$ at DAFNE are well suited to see this resonance if one selects forward going neutrons in the center of mass, which reduce the contribution of single scattering and stress the contribution of the double scattering where the signal of the resonance appears clearer. We take advantage to report briefly on a recent work in which in addition to the $\bar{K}NN$ system with total spin S=0, we find a less bound state (although with equally large width) with S=1, like in the $K^{-} d$ reported in the first part.}

%

\section{Introduction}

In a recent paper \cite{sekideu} we reported on the interest of the $K^{-} d \to \pi \Sigma n$ reaction, which was measured in \cite{expe}. The idea is that the $K^{-}$ scatters with a neutron, loses energy and can interact with the proton to produce the $\Lambda(1405)$. In this way the production of the resonance is induced by a $K^{-}$ and this guarantees that the state excited if mostly the one appearing at higher energy and narrow, out of the two $\Lambda(1405)$ states found in \cite{cola}. One of the conditions for the success of the experiment was to use kaons in flight. The reason is that the process that shows clearly the resonance peak is the double scattering. If the kaon is away from threshold, the dominant one body scattering is far away of the resonance region and peaks at higher energies. At lower energies of the kaon, the peak of the one body collision appears close to threshold and blurs the signal of the $\Lambda(1405)$ coming from double scattering.  The DAFNE conditions, where the kaons come from the decay of the $\phi$, produce low energy kaons. Yet, we have found in \cite{dafne} that, 
in spite of the low momenta of the kaons, one can still see the good signal for the resonance, but on the condition that neutrons are detected simultaneously in the forward direction in the CM, which drastically reduces the background from single scattering. Since we are working with the $K^{-} d$ reaction, we also report on recent results from \cite{melahat}, where in addition to the $\bar{K}NN$ system with total spin S=0, reported in several theoretical works \cite{gal,sato,hyodo}, we find a less bound state (although with equally large width) with S=1. 

\section{The formalism}

The ${\cal T}$  matrix for the $K^{-} d \to \pi \Sigma n$ reaction is given by the sum of the contribution of the three diagrams of  fig.~\ref{fig2}, ${\cal T} =  {\cal T}_{1} + {\cal T}_{2} + {\cal T}_{3}$,
where the different amplitudes are given by 

\begin{equation}
  {\cal T}_{1} = T_{K^{-}p \rightarrow \pi \Sigma}(M_{\pi\Sigma}) \, \varphi(\vec p_{n} - \frac{\vec p_{d}}{2}). \label{eq:T1}
\end{equation}

\begin{eqnarray}
  {\cal T}_{2}& =&  T_{K^{-}p \rightarrow \pi \Sigma}(M_{\pi\Sigma}) 
 \int \frac{d^{3}q}{(2\pi)^{3}} \frac{\tilde \varphi (\vec q+\vec p_{n}-\vec k - \frac{\vec p_{d}}{2})}{q^{2}-m_{K}^{2} + i\epsilon}  
 T_{K^{-}n \rightarrow K^{-}n}(W_{1}) \ . \label{eq:T2}
\end{eqnarray}

\begin{eqnarray}
  {\cal T}_{3}& =& - T_{\bar K^{0}n \rightarrow \pi \Sigma}(M_{\pi\Sigma}) 
  \int \frac{d^{3}q}{(2\pi)^{3}} \frac{\tilde \varphi (\vec q+\vec p_{n}-\vec k - \frac{\vec p_{d}}{2})}{q^{2}-m_{K}^{2} + i\epsilon}
  T_{K^{-}p \rightarrow \bar K^{0}n}(W_{1})  \ . \label{eq:T3}
\end{eqnarray}
with $\varphi(\vec p_{n} - \frac{\vec p_{d}}{2})$ the deuteron wave function in momentum space and 
\begin{eqnarray}
   q^{0} &=& M_{N} + k^{0} - p_{n}^{0}\ , \\
   W_{1} &=& \sqrt{(q^{0}+p_{n}^{0})^{2}-(\vec q + \vec p_{n})^{2}} \ .
\end{eqnarray}
For $q^0$ we have assumed that the deuteron at rest has energy 
$2M_N -B$, and we have taken half of it for one nucleon, 
neglecting the small binding energy. The variable $W_1$ depends, 
however, on the running $\vec{q}$ variable.

\begin{figure}
\begin{center}
\centerline{\includegraphics[width=8.5cm]{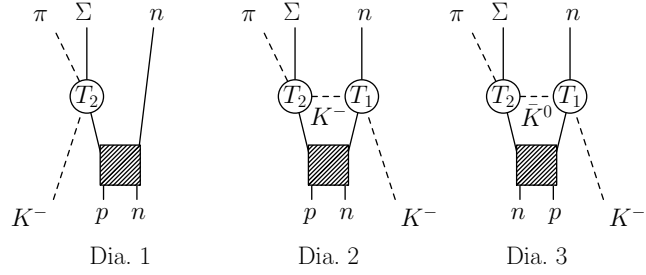}}
\caption{Diagrams for the calculation of the $K^{-}d \to \pi\Sigma n$ reaction.
$T_{1}$ and $T_{2}$ denote the scattering amplitudes for $\bar KN \to \bar KN$
and $\bar K N \to \pi \Sigma$, respectively.  \label{fig2}}
\end{center}
\end{figure}

 In fig. \ref{fig4} we show the results for the cross sections for different intervals of neutron angle with respect to the original $K^-$ in the center of mass frame. We find that when the angles are chosen small the signal of the resonance appears neatly, because the condition of the neutron detection in coincidence has drastically reduced the contribution form the single collision. We have checked that even with the small signal that we find the cross sections are sufficiently large to be measure at DAFNE, collecting data for a few months while other experiments are running at the same time \cite{nevio}
\begin{figure}
\begin{center}
%
\includegraphics[width=15cm]{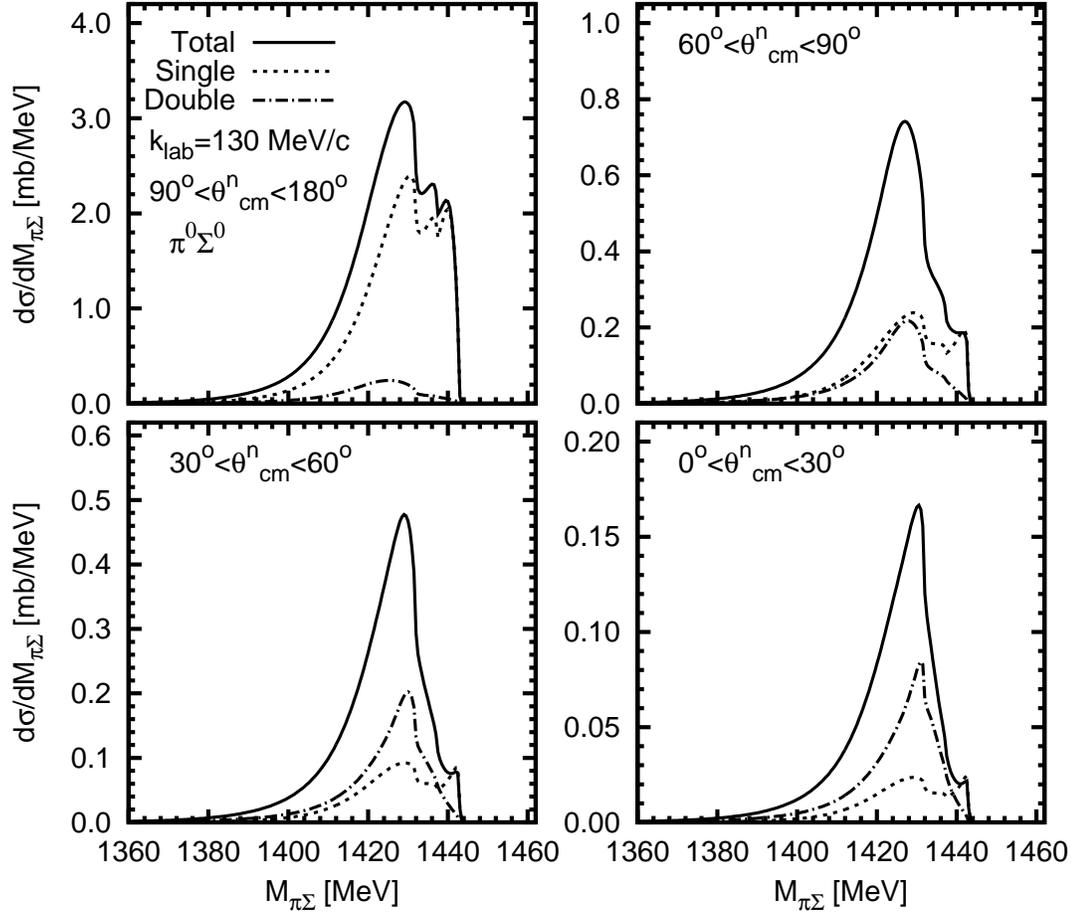}
%
\caption{$\pi\Sigma$ invariant mass spectra of $K^{-}d \to \pi^{0}\Sigma^{0}n$
with 130 MeV/c of incident $K^{-}$ momentum imposing angular cuts 
for the emitted neutron with respect to the incident $K^{-}$ in the CM frame, 
$90^{\circ}<\theta^{n}_{\rm c.m.}<180^{\circ}$ (up-left),
$60^{\circ}<\theta^{n}_{\rm c.m.}<90^{\circ}$ (up-right),
$30^{\circ}<\theta^{n}_{\rm c.m.}<60^{\circ}$ (down-left) and
$0^{\circ}<\theta^{n}_{\rm c.m.}<30^{\circ}$ (down-right).
In each panel, the solid line denotes 
the total contributions of the three diagrams, while the dotted and 
dash-dotted lines show the calculations 
from the single and double
scatterings, respectively. 
}
\label{fig4}
\end{center}
\end{figure}

\section{The $\bar{K} NN$ system in S=1}

In several papers it has been reported that there is a bound state of $\bar{K} NN$ with S=0 \cite{gal,sato,hyodo}. They use different techniques to get the state, Faddeev equations in  \cite{gal,sato} and a variational calculation in 
\cite{hyodo}. In all cases they looked for the most bound state \cite{hyodo} or explicitly for the S=0 state, for which one has reasons that it is the most bound state. The S=1 system was not looked at. Ina recent paper \cite{melahat}, this system has been studied, together with the S=0 one, by means of the Fixed Center Approximation to the Faddeev equations, and it has been found that indeed, the S=0 state is the most bound. Upon changing the size of the NN system to agree with the results of \cite{hyodo}, we find that the results for the S=0 system are very similar to those of \cite{hyodo}, which uses the same input from the chiral Lagrangians, with a binding energy around 40 MeV, and a width without consideration of $K^-$ absorption of 50 MeV. However, we also find a system bound with S=1, like one would have for $K^- d$ in s-wave, as we have studied in the first part. However, this state is less bound than the one for S=0, only about 27 MeV and a similar width. 
   Interesting as these results re, we should keep in mind that the experimental finding of such states, is certainly difficult, since the width is larger than the binding energy, Our finding now that there is a lower sate with a different spin, can only complicate the experimental search further, c
since in many possible experiments the contribution from the two states would produce overlapping states with also a large width, which could blur any signal of a possible bound state.



}  


\end{document}